\documentclass[prl,onecolumn,superscriptaddress,amsmath,amssymb]{revtex4}

\usepackage{graphicx}% Include figure files
\usepackage{dcolumn}% Align table columns on decimal point
\usepackage{bm}% bold math
\usepackage{hyperref}% add hypertext capabilities
\usepackage{amsmath}
\usepackage{amssymb}
\usepackage{subfigure}
\usepackage{epstopdf}
\usepackage{color}
\usepackage{wasysym}

\begin{document}

\begin{abstract}
We consider the persistent currents induced by an artificial gauge field applied to interacting ultra-cold bosonic atoms in a tight ring trap. Using both analytical and numerical methods, we study the scaling of the persistent current amplitude with the size of the ring. In the strongly interacting regime we find a power-law scaling, in good agreement with the predictions of the Luttinger-liquid theory. By exploring all interaction regimes we find that the scaling is optimal, i.e. the current amplitude decreases slower with the system size, at intermediate interactions. 
\end{abstract}

\title{Optimal scaling of persistent currents for interacting bosons on a ring}

\author{Marco Cominotti}
\affiliation{Universit\'e Grenoble Alpes, LPMMC, F-38000 Grenoble, France}
\affiliation{CNRS, LPMMC, F-38000 Grenoble, France}

\author{Matteo Rizzi}
\affiliation{Institut f\"ur Physik, Johannes Gutenberg-Universit\"at Mainz,Staudingerweg 7, D-55099 Mainz, Germany}

\author{Davide Rossini}
\affiliation{NEST, Scuola Normale Superiore and Istituto Nanoscienze-CNR, I-56126 Pisa, Italy}

\author{Davit Aghamalyan}
\affiliation{Centre for Quantum Technologies, National University of Singapore, Singapore 117543}

\author{Luigi Amico}
\affiliation{CNR-MATIS-IMM $\&$ Dipartimento di Fisica e Astronomia,  I-95127 Catania, Italy}
\affiliation{Centre for Quantum Technologies, National University of Singapore, Singapore 117543}

\author{Leon C. Kwek}
\affiliation{Centre for Quantum Technologies, National University of Singapore, Singapore 117543}
\affiliation{National Institute of Education and Institute of Advanced Studies, Nanyang Technological University, Singapore 637616}

\author{Frank Hekking}
\affiliation{Universit\'e Grenoble Alpes, LPMMC, F-38000 Grenoble, France}
\affiliation{CNRS, LPMMC, F-38000 Grenoble, France}

\author{Anna Minguzzi}
\affiliation{Universit\'e Grenoble Alpes, LPMMC, F-38000 Grenoble, France}
\affiliation{CNRS, LPMMC, F-38000 Grenoble, France}

\maketitle

\section{1. Introduction}

In a quantum fluid confined on a toroidal geometry, the persistent current phenomenon is the periodic particle-current response to the flux of an applied $U(1)$ gauge potential. This is a manifestation of the Aharonov-Bohm effect at the many-body level, indicating that the phase coherence length of the fluid extends to the whole system. Persistent currents were first observed in electronic systems subjected to a magnetic field, such as bulk superconductors~\cite{Deaver,byers,onsager}, and more recently also in normal resistive metal rings~\cite{Levy,mailly,bluhm,harris}. The observation of persistent currents in electronic systems is challenging because of their high sensitivity to the electromagnetic environment~\cite{kravtsov}, and to the microscopic disorder present within the ring~\cite{schwab}. Interaction effects are instead negligible, and models based on single particle diffusion are in good agreement with experiments~\cite{harris}.

Ultracold atomic gases are model quantum fluids, highly tunable and extremely clean. Interactions, dimensionality, confining potentials and even disorder can be adjusted at will, giving rise to a versatile and rich platform to investigate quantum many-body phenomena. In recent experiments, an atomic circuit based on a flowing toroidal Bose-Einstein condensate in a micrometric ring trap has been realized~\cite{gupta,ryu,beattie,ramanathan,wright}. For these neutral atomic samples the presence of a gauge potential can be engineered by stirring the gas at constant velocity with a localized barrier potential, realized using well-focused repulsive laser beams~\cite{ramanathan,wright}. %, or by directly imparting geometric phases to the atoms via suitably designed laser fields~\cite{dalibard}.
Current states have been achieved in this way and probed via the time-of-flight expansion. These experiments are important in view of the realization of current-based devices, such as atomtronic SQUID analogues or flux qubits~\cite{flux-qubit,solenov_qubit,hallwood,rey,amico}.

In a recent work we have considered an interacting one-dimensional (1D) Bose gas confined in a ring trap,
subjected to such an artificial $U(1)$ gauge potential~\cite{marco}.
There, we have studied how the persistent current amplitude depends, at fixed ring length $L$, on the barrier height and on the interaction strength, finding an optimal regime at intermediate interaction strength, where the current amplitude is maximal, due to the interplay between classical screening and quantum fluctuations. In the present work, we focus on
the mesoscopic nature of the persistent current, which vanishes for macroscopic system sizes (i.e., $L \rightarrow \infty$).
Interestingly, we find that the scaling behavior of the current amplitude is again optimal at intermediate interaction strength, where the current amplitude decreases slower with the system size.

\section{2. Model and Methods}
\label{section:Model}

We consider a system of $N$ bosons of mass $m$, subjected to contact interaction, accounting for s-wave scattering, and loaded into a 1D ring of length $L$, with density $n=N/L$. The position along the ring is parametrized by the angular coordinate $\theta \in [0, 2\pi)$. The ring contains a localized barrier potential, modeled as a delta function, and is threaded by an artificial (dimensionless) `magnetic flux' $\Omega$. The corresponding Hamiltonian reads
\begin{equation}
  \mathcal{H} \! = \ E_0 \, \sum_{j=1}^{N}
 \! \left[ \left( \!\! -i\frac{\partial}{\partial \theta_{j}} - \Omega \! \right)^{\!\! 2} \!
  + \frac{L}{2 \pi a} \Lambda \, \delta(\theta_j) +\frac{N\gamma}{2\pi} \! \sum_{l=1}^{N}\delta(\theta_{l}-\theta_{j}) \right] \,,
  \label{eq:hamiltonian}
\end{equation}
where we adopted the zero-point single particle kinetic energy $E_{0}=2\pi^{2}\hbar^{2}/mL^2$ as natural energy unit,
$\Lambda$ is the strength of the barrier, $a$ is a length unit which will be specified below, and $\gamma=mg/\hbar^{2}n$ is the dimensionless interaction parameter, with $g$ the contact interaction coupling.
The spatially-averaged particle current $I(\Omega)$ has been obtained, at zero temperature and in the stationary regime, from the ground-state energy $E(\Omega)$ via the thermodynamic relation $I(\Omega)=-(1/2\pi\hbar)\partial E(\Omega)/\partial \Omega$~\cite{Bloch1970}.

In the absence of the barrier and for any interaction strength, the ground-state energy of Hamiltonian~\eqref{eq:hamiltonian} is a series of parabolas, $E(J,\Omega) = N E_0 (\Omega - J)^2 + E_\mathrm{int}$, where $E_\mathrm{int}$ is the $\Omega$-independent interaction energy, corresponding to states of well-defined angular momentum and circulation $J$, periodically shifted by a Galilean transformation in $\Omega$ with period 1, and intersecting at the frustration points $\Omega_j = (2 j +1)/2$. The corresponding persistent current is a perfect sawtooth of amplitude $I_{0} = N E_0 / \pi \hbar= 2\pi \hbar \, n / m L$~\cite{leggett,loss}.  We notice that the amplitude of the persistent current vanishes in the thermodynamic limit ($N\rightarrow\infty$, $L\rightarrow\infty$, at fixed $n=N/L$), as is well known~\cite{harris,riedel}.

\begin{figure}
\centering
\resizebox{0.35\columnwidth}{!}{\includegraphics{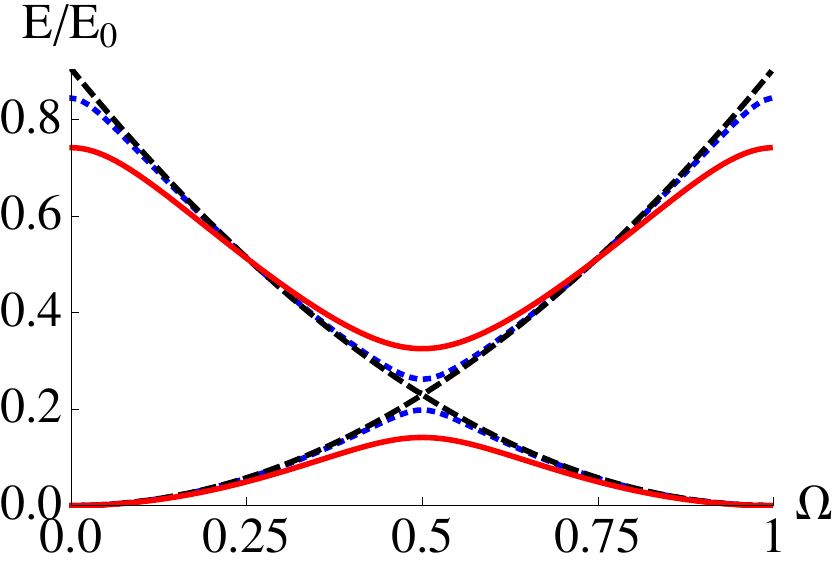} }
\resizebox{0.398\columnwidth}{!}{\includegraphics{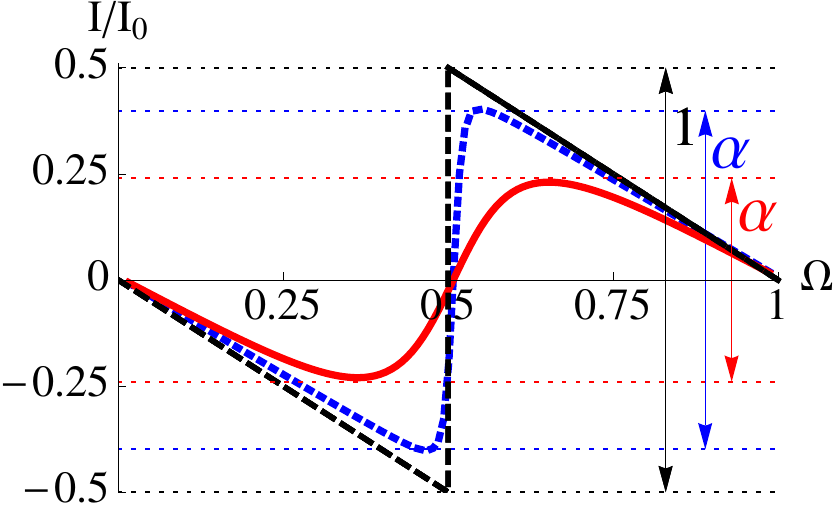} }
  \caption{Energy spectrum and persistent current in the first rotational Brillouin zone, in the TG limit, for density $n=0.25/a$, and zero barrier (black dashed line), barrier strength $\Lambda=0.1$ (blue dashed line), and $\Lambda=1$ (red solid line).}
  \label{sawtooth}
\end{figure}

In the presence of a barrier that breaks the rotational symmetry, a gap opens in the many-body energy spectrum at the frustration points, mixing states that differ by one quantum of circulation. The corresponding persistent current is smeared, with a shape depending on barrier and interaction strengths~\cite{hekking,marco}. In Fig.~\ref{sawtooth} we show an example of the energy spectrum and of the corresponding particle current as a function of the flux $\Omega$.
In the following, we study the scaling of the current amplitude $\alpha=(I_{\max}-I_{\min})/I_{0} = 2 I_{\max} / I_0$ with the system size, keeping fixed the density $n$, in the presence of barrier and interactions.

In order to solve the many-body problem associated with Hamiltonian (\ref{eq:hamiltonian}) in all the regimes of interaction $\gamma$ and barrier strength $\Lambda$, we have resorted to a combination of numerical and analytical, exact and approximate, techniques, for which more details are given in Ref.~\cite{marco}.
At arbitrary barrier strength and for intermediate-to-strong interactions we have adopted a numerical technique based on the density matrix renormalization group (DMRG), in which space has been discretized in $M$ lattice sites of spacing $a=L/M$, that we take as the length unit, and Hamiltonian (\ref{eq:hamiltonian}) has been mapped onto a 1D Bose-Hubbard (BH) model: 
$\mathcal{H}_{\rm \scriptstyle BH} =\sum_{j=1}^{M} [-t( e^{-i\frac{2\pi\Omega}{M}} b^\dagger_j b_{j+1}  +  {\rm H.c.} )+ (U/2) n_j (n_j - 1) + \Lambda\delta_{j,1} n_{j}]$,
where $b^\dagger_j$ ($b_j$) are bosonic creation (annihilation) operators at site $j$, $n_j=b^\dagger_jb_j$,
$t$ is the tunnel energy on adjacent sites, $U$ is the on-site interaction energy, and $2\pi\Omega / M$ is the Peierls phase twist induced by the gauge field~\cite{comment0}.
In the two opposite limiting cases of non interacting (NI) and infinitely interacting Tonks-Girardeau (TG) gas, the many-body problem reduces to a single-particle one and is therefore solved exactly~\cite{girardeau}.
At weak interactions the bosonic fluid is described within a mean-field approximation via the Gross-Pitaevskii (GP) equation. Its ground state is a dark soliton pinned at the position of the barrier~\cite{marco}.

Finally, in order to give a theoretical interpretation to the numerical results at strong interaction, we use the Luttinger liquid (LL) theory, a low-energy quantum hydrodynamics description of the bosonic fluid, which holds in the intermediate- to strong-interaction regime. We treat perturbatively the barrier contribution to the Hamiltonian, in the two opposite cases of small and large barrier strength~\cite{marco}. For small barrier strength we obtain the persistent current
$I(\Omega)=-I_{0} \, \delta\Omega ( 1-{\pi/\sqrt{ (2 \pi \delta\Omega)^{2}+(\Lambda_{\rm eff}L/2\pi a)^{2}}})$,
where $\delta\Omega=\Omega-1/2$, $\Lambda_{\rm eff}= \Lambda (d/L)^K$ is the effective barrier strength renormalized by the density quantum fluctuations, and $d$ is the short-distance cutoff of the LL theory. $K$ is the Luttinger parameter, whose dependence on the interaction strength is known for the Lieb-Liniger model from Bethe ansatz~\cite{cazalilla,comment}. $K=1$ in the TG limit, and $K$ increases at decreasing interaction tending to infinity towards the NI one.
The amplitude of the current is then given by $\alpha=1-(3/2)(\Lambda_{\rm eff} L/ 2 \pi^{2} a)^{2/3}$, and hence we obtain the scaling

\begin{equation}
1-\alpha \sim L^{(2/3)(1-K)}\,.
\label{eq:scaling}
\end{equation}
In the opposite regime of large barrier strength, we obtain the persistent current $I(\Omega)=-(2t_{\rm eff} n/ \hbar)\sin(2\pi\Omega)$, where $t_{\rm eff}=t(d/L)^{1/K}$ is the effective tunneling amplitude across the barrier, renormalized by the quantum fluctuations of the phase. The tunneling amplitude $t$ across the barrier, which is small in the large-barrier limit, is related to the barrier strength via $t/L \sim \left( \hbar\omega_{c}\right)^{1+K} {(L/a \Lambda)^{K}}$, with cut-off energy $\hbar\omega_c\sim NE_0 \sim {n L^{-1}}$. Therefore the current amplitude for large barrier scales as
\begin{equation}
\alpha \sim L^{1-1/K}\,.
\label{eq:scaling2}
\end{equation}
In the next section, we will compare the predictions of Eqs.~(\ref{eq:scaling}) and~(\ref{eq:scaling2}) with the numerical data,
showing that they display a good agreement.

\begin{figure}[b!]
  \centering
  \resizebox{0.297\columnwidth}{!}{ \includegraphics{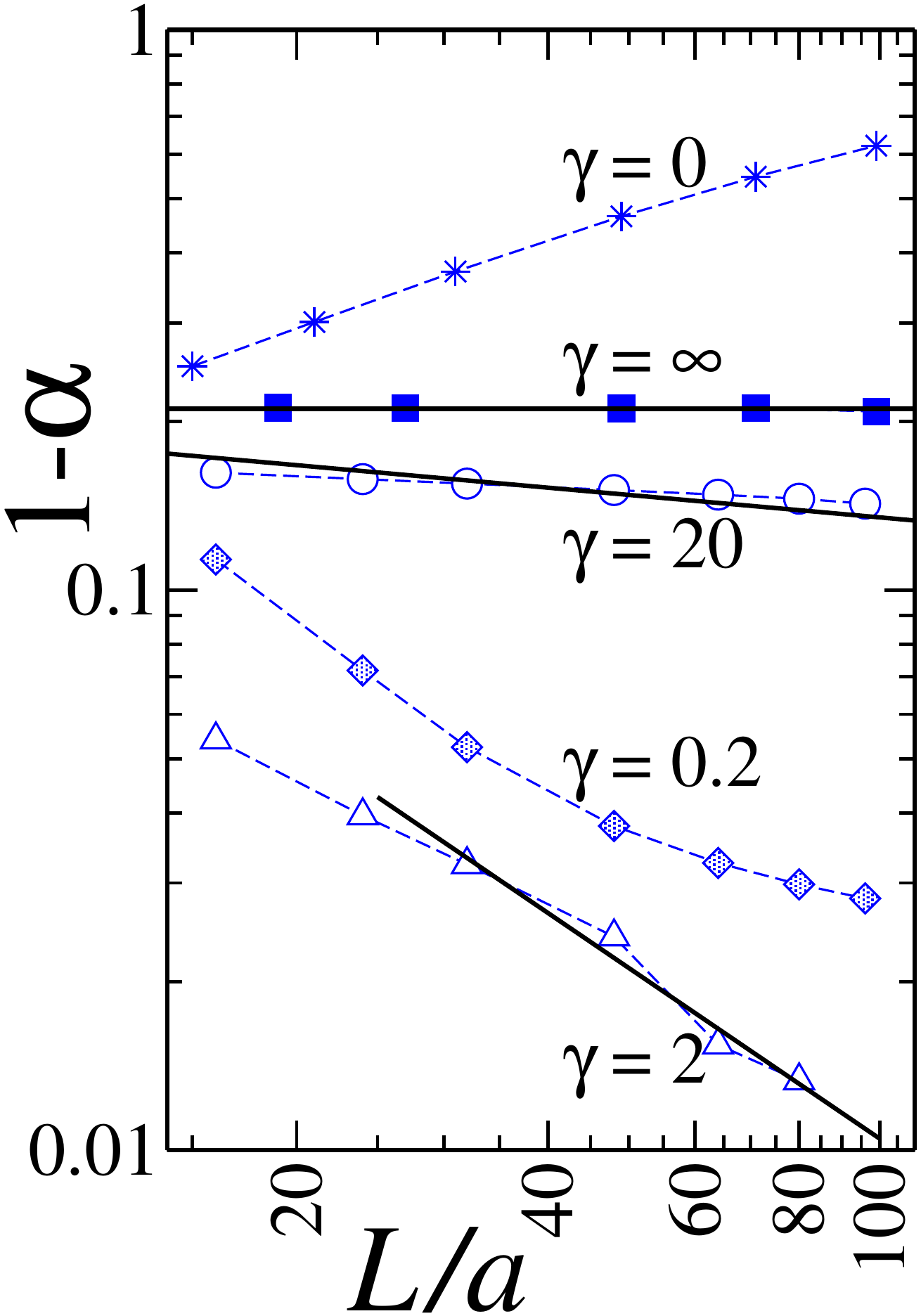} }
  \resizebox{0.304\columnwidth}{!}{ \includegraphics{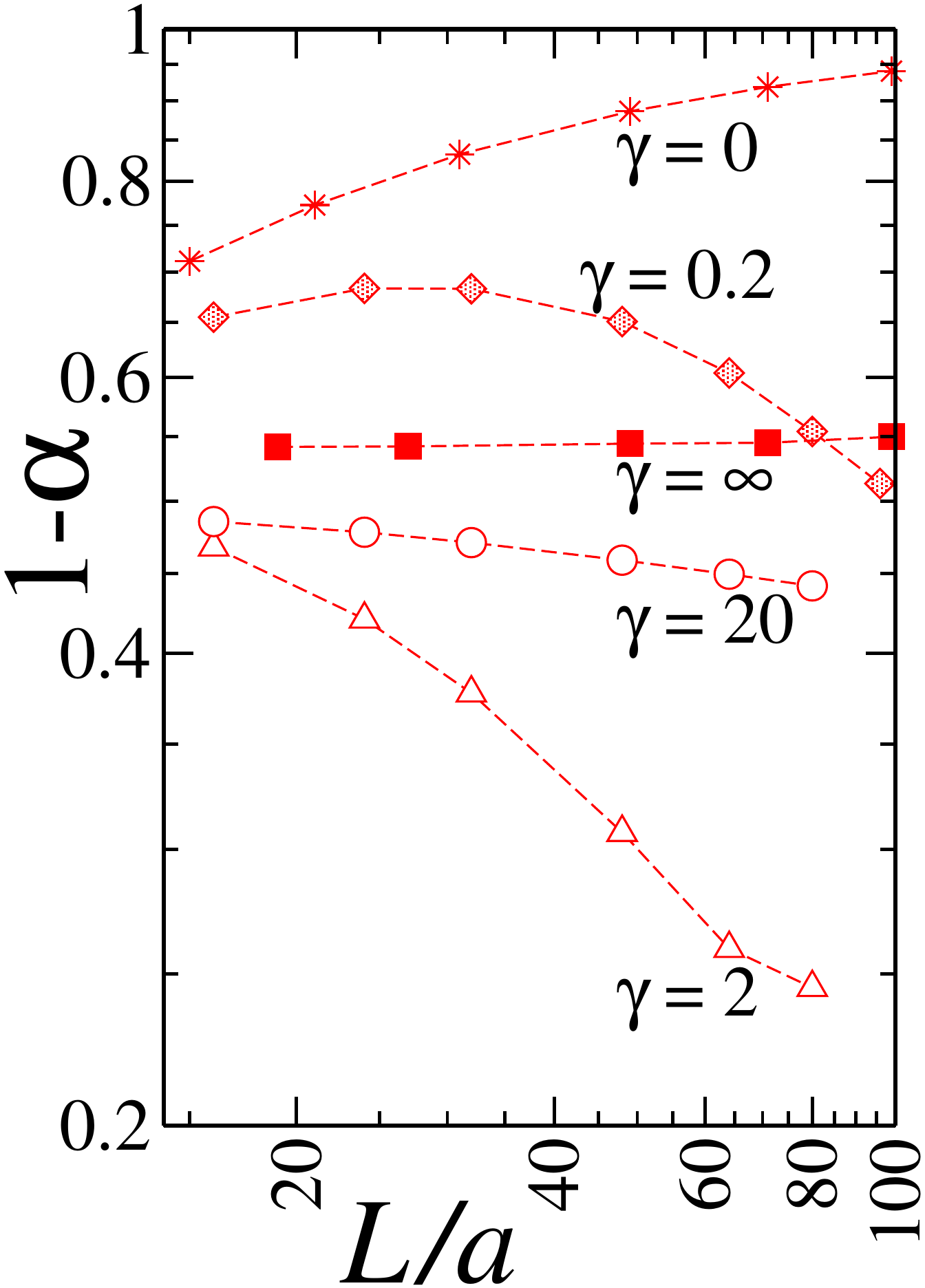} }
  \resizebox{0.297\columnwidth}{!}{ \includegraphics{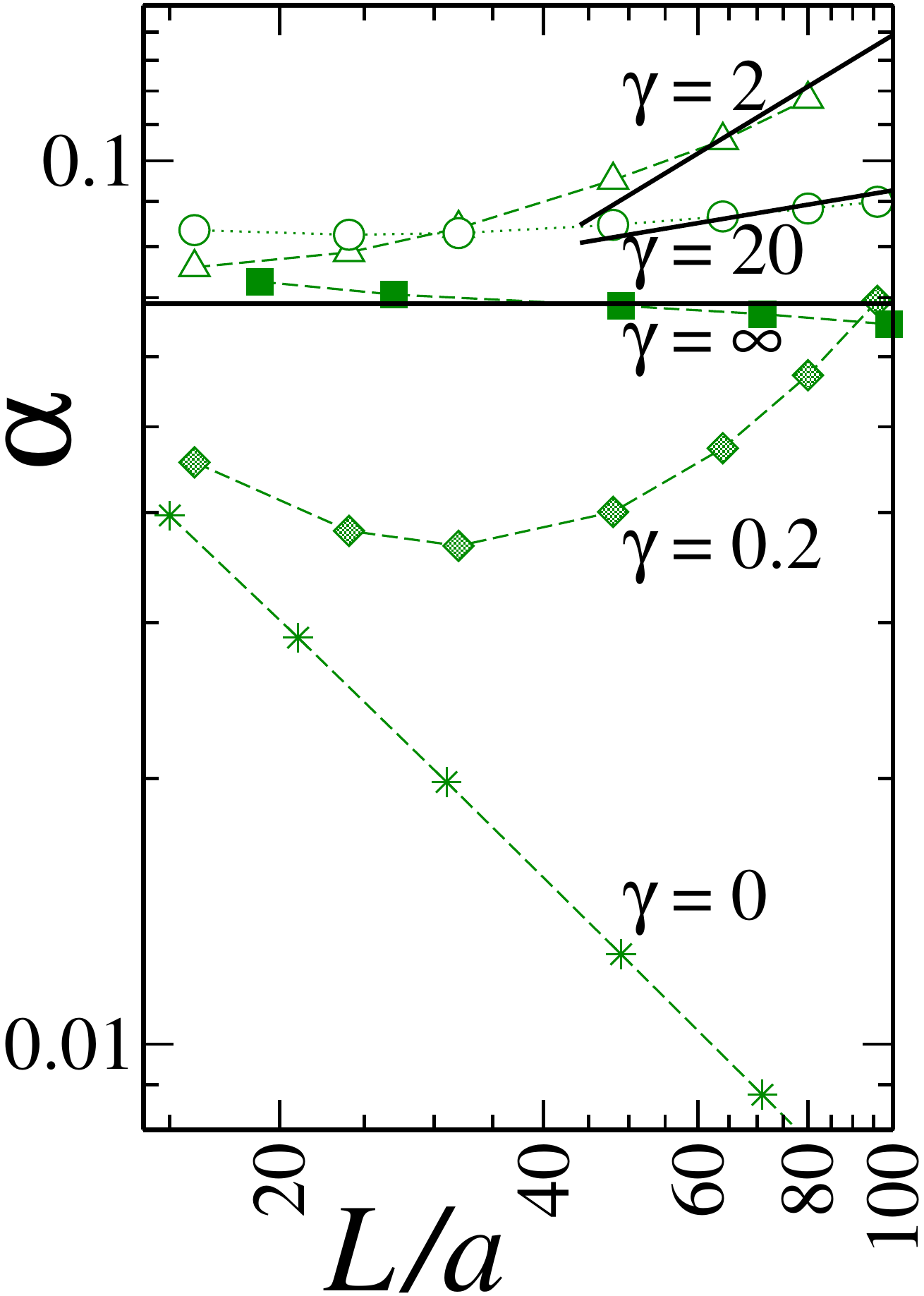} }
  \caption{Log-Log plots of the persistent current amplitude versus system size $L/a$, at fixed density $n=0.25/a$ at various values of the interactions $\gamma=\infty$ (TG ($\blacksquare$)), $20$ (DMRG ($\bigcirc$)), $2$ (DMRG ($\triangle$)), $0.2$ (GP ($\diamondsuit$)), $0$ (NI ($*$)).
 For weak barrier strength $\Lambda=0.1$ (blue) and $1$ (red), we plot $1-\alpha$, with faster decaying curves being more favorable for the current, while for large barrier strength $\Lambda=10$ (green) we more conveniently show $\alpha$, with reverted meaning of the scaling behavior.
The solid black lines show the predictions of the LL, Eq.~(\ref{eq:scaling}) for weak barrier (first panel) and~(\ref{eq:scaling2}) for strong barrier (third panel), with LL parameter $K|_{\gamma=\infty}=1.00$, $K|_{\gamma=20}\simeq1.20$ and $K|_{\gamma=2}\simeq2.52$, as extracted from the asymptotic expansions of $K(\gamma)$~\cite{cazalilla}.}
  \label{scaling}
\end{figure}

\section{3. Scaling of persistent currents with system size}

In Fig.~\ref{scaling} we show the scaling of the persistent current amplitude $\alpha$ with the system size, at fixed density $n = 0.25/a$, for various values of barrier and interaction strengths. We observe that for all values of the barrier the scaling strongly depends on the interaction strength. In the NI regime the scaling is unfavorable, because increasing the system size the current amplitude vanishes even faster than $n L^{-1}$, the overall scaling factor encoded in the current unit $I_0$. At increasing interactions, we observe instead that the scaling gets more favorable, because the decay of the current amplitude with the system size is, in part, compensated by many-body effects. In particular, we observe that there is an optimal regime, at intermediate interactions ($\gamma\simeq 2$), for which the scaling is the most favorable, in all the regimes of barrier height. In the first and third panels of Fig.~\ref{scaling} we notice that, at large enough interactions, the current amplitude obtained numerically scales as a power law, in good agreement with the Luttinger-liquid expressions (\ref{eq:scaling}) and~(\ref{eq:scaling2}) respectively.

The presence of an optimal regime can be understood in terms of screening of the barrier. 
In the semi-classical GP regime, this is determined by the healing length $\xi=\hbar/\sqrt{2mgn}$, which gets smaller and smaller while increasing the interactions. This means that the barrier gets more and more invisible to the fluid, and we can observe a gain in the scaling rate.
At stronger interaction, beyond the regime of applicability of the GP equation, quantum fluctuations become crucial, especially in a 1D system, as can be understood from the LL description. 
Density fluctuations, which screen the barrier as well, are stronger at intermediate interactions ($K\gg1$) and get suppressed at larger ones ($K\sim1$), where phase fluctuations, which spoil the coherence, dominate~\cite{marco,cazalilla}. 
Furthermore, their contribution in either sense grows with the system size, as one can see from the expressions of $U_{\rm eff}$ and $t_{\rm eff}$ in section \ref{section:Model}2.

\section{4. Conclusions}
 Persistent currents are a mesoscopic phenomenon: their amplitude vanishes in the thermodynamic limit. In this work we have studied the scaling of the persistent currents with system size for the case of one-dimensional interacting bosons. We have found that the scaling depends on the interaction strength, and that the persistent current amplitude decreases slower at intermediate interactions than at very large or very small ones. This non-monotonic effect is due to the combination of the effects of interaction and quantum fluctuations. 
Our result is important in view of the forthcoming experimental realizations, where the best regime for observing the largest possible current signal should be found from the trade-off between realizable small system size and interaction strength.

\section*{Acknowledgments}
This work is supported by the ERC Handy-Q grant
N.258608,  Institut universitaire de France,  Italian MIUR through FIRB
Project RBFR12NLNA. Numerical simulations were performed on the MOGON cluster in Mainz.


\begin{thebibliography}{}
\bibitem{comment0} 
The BH parameters are linked to the continuous model ones by: $\hbar^{2}/2m=ta^2$, and $\gamma=UM/2tN$.
\bibitem{comment} 
 We have also checked that the same value of $K$ is obtained by fitting the DMRG one-body density matrix with the LL power-law expression $\sim|\sin(\pi x/L)|^{-1/2K}$, thereby validating the use of the BH model for our low-filling regime to describe the continuous one.
\end{thebibliography}
\end{document}